\documentclass[multphys,vecphys]{svmult}

\usepackage{makeidx}     
\usepackage{graphicx}    
\usepackage{multicol}    

\makeindex             

\begin{document}

\title*{Observational Properties of Type II Plateau Supernovae}
\author{A. Pastorello\inst{1,}\inst{2}, 
M. Ramina\inst{1,}\inst{2}, L. Zampieri\inst{2},\\ 
H. Navasardyan\inst{2}, M. Salvo\inst{3}, M. Fiaschi\inst{1}}
\authorrunning{A. Pastorello et al.} 
\institute{Dipartimento di Astronomia, Universit\`a di Padova, 
Vicolo dell' Osservatorio 2, I-35122 Padova, Italy
\and INAF - Osservatorio Astronomico di Padova, Vicolo 
dell' Osservatorio 5, I-35122 Padova, Italy 
\and Australian National University - Mount Stromlo Observatory,
Cotter Road, Weston ACT 2611, Australia
}
%
%
\maketitle

We present spectroscopic and photometric data of a 
 sample of Type II plateau Supernovae, covering a wide range of 
 properties, from the $^{56}$Ni rich,
 high luminosity events (e.g. SN~1992am) to 
 the low--luminosity, $^{56}$Ni poor SNe (e.g. SN 1997D). 
 We provide an observational framework to analyze correlations 
 among observational data, physical parameters and 
 progenitors characteristics of Type II Supernovae. 

\section{The Sample of SNe II--P}
Type II plateau Supernovae (SNe II--P) are considered a heterogeneous group of core--collapse events 
sharing a very wide range of physical properties. 
Despite their variety of observational parameters 
(e.g. early-- and late--time luminosity, expansion velocity, continuum 
temperature), recent studies highlight tight correlations among their
physical parameters \cite{hamu03,nady03}.
However, in these works the low--luminosity tail of the SNe II--P distribution
was poorly sampled. Zampieri et al. [these Proceedings] have recently investigated such 
correlations, including also low--luminosity events.
Our sample was selected in such a way to cover a large
range in luminosity and line velocity, preferably among SNe II--P discovered
at very early stages. We selected a few well studied SNe from literature and 
unpublished data from the Padova--Asiago SN Archive.
Most of them have long--duration plateaux,  but also events with relatively short plateaux
(SNe~1992H and 1995ad) and spectroscopic evolution of a normal SN II--P, were considered. 
SNe from our archive have been observed either in spectroscopy and
photometry from a few days after their discovery
to the nebular phase, when the main output of energy comes from the
radioactive decays.
The sample includes 6 Ni poor ($<$ 10$^{-2} M_\odot$) SNe,
5 intermediate Ni mass (1--5$\times$10$^{-2} M_\odot$) SNe and 5 more rich
Ni mass ($>$ 7$\times$10$^{-2} M_\odot$) events.

\begin{table}
\centering
\caption{Main data of the selected SNe II--P}
\label{tab1}       
\begin{tabular}{ccccc|ccccc}
\hline\noalign{\smallskip}SN & $\mu$ & A$_{V}$ & t$_0$ (JD) & ref.($\star$) & SN & $\mu$ & A$_{V}$ & t$_0$ (JD) & ref.($\star$) \\
\noalign{\smallskip}\hline\noalign{\smallskip}
1969L & 29.84 & 0.20 & 2440550.5 & \cite{ciat71} & 1996an & 31.50 & 0.16 & 2450222 & \cite{pasto03tesi} \\
1987A & 18.49 & 0.60 & 2446849.82 & SAAO & 1997D & 31.29 & 0.07 & 2450361 & \cite{tura98},\cite{bene01} \\
1992H & 32.48 & 0.33 & 2448661 & \cite{cloc96} & 1999br & 31.19 & 0.08 & 2451278 & \cite{hamu01},\cite{pasto03} \\
1992am & 36.74 & 0.44 & 2448799 & \cite{schm94} & 1999em & 29.47 & 0.31 & 2451476 & \cite{abou03}, \cite{leon02} \\
1992ba & 30.91 & 0.19 & 2448883.2 & \cite{hamu01},\cite{pasto03tesi}& 1999eu & 31.08 & 0.09 & 2451394 & \cite{pasto03} \\
1994N & 33.34 & 0.13 & 2449451 & \cite{pasto03} & 2001dc & 32.85 & 1.28 & 2452056 & \cite{pasto03} \\
1995ad & 32.02 & 0.11 & 2449981 & \cite{pasto03tesi} & 2002gd & 33.09 & 0.22 & 2452552 & \cite{pasto03tesi} \\
1996W & 31.95 & 0.70 & 2450180 & \cite{pasto03tesi} & 2003Z & 31.93 & 0.13 & 2452665 & \cite{pasto03tesi} \\
\noalign{\smallskip}\hline 
\end{tabular}\\
($\star$) reference for  spectro--photometric data
\end{table}

In Tab. \ref{tab1} we list the main data about the selected SNe (see also
Ramina, Laurea Thesis, 2003, unpublished and references therein). When the
distance modulus $\mu$ is obtained from the host galaxy recession velocity,
H$_0$ has been assumed to be equal to 65
km s$^{-1}$ Mpc$^{-1}$. The total extinction reported in Tab. \ref{tab1} 
is the sum of the host galaxy reddening plus the Galactic contribution,
from Schlegel et al. \cite{schl98}. More details on the estimated distances and
interstellar extinction are in Ramina [Laurea Thesis, 2003, unpublished].  
\vspace{-0.5cm}
  
\begin{figure}
\centering
\includegraphics[height=6.7cm,width=8cm]{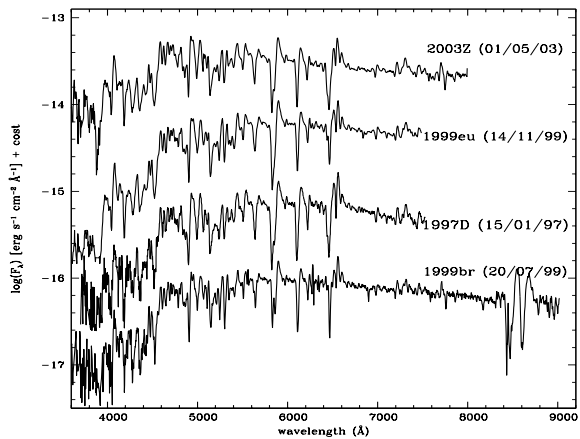}
\caption{Spectra of low--luminosity SNe II--P at $\sim$100 days after the explosion.}
\label{fig1}       
\end{figure}

\vspace{-0.6cm}

\subsection{Faint SNe II--P}
\label{faint}
\begin{figure}
\hspace{-.5cm}
\includegraphics[height=5.9cm]{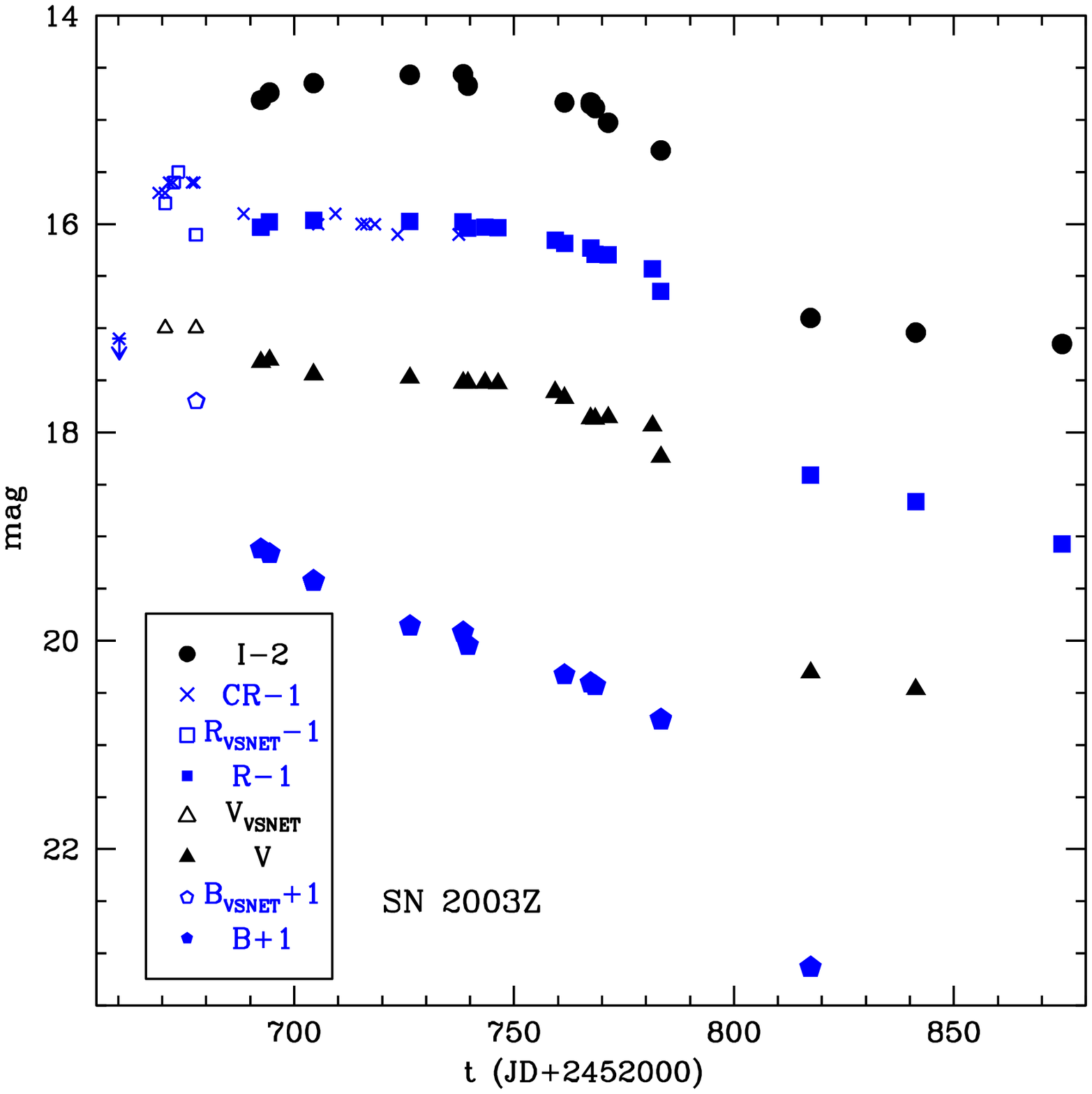}
\hspace{-0.5cm}
\includegraphics[height=5.8cm,width=6.4cm]{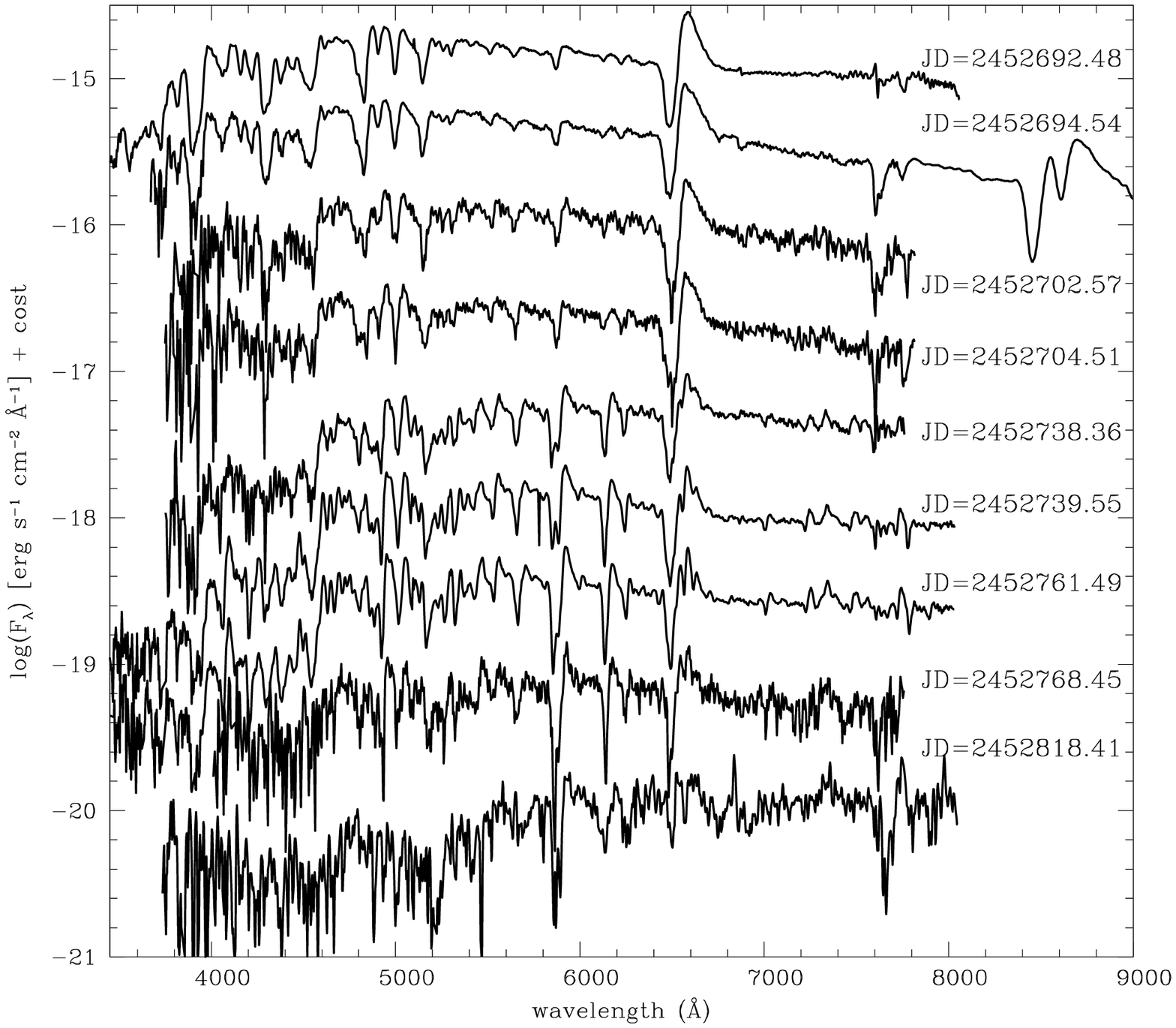}
\caption{Photometric and spectroscopic evolution of the low--luminosity SN~2003Z.
Unfiltered measurements and VSNET {(\sl http://vsnet.kusastro.kyoto-u.ac.jp/vsnet/)} data are also reported.}
\label{fig3}     
\end{figure}

SN~1997D \cite{tura98,bene01} is the prototype of a homogeneous group of
CC--SNe with unique observational properties. 
The light curves, showing flat plateaux lasting $\sim$ 90--110 days, 
are underluminous at all epochs,
and their spectra, redder than ``typical'' SNe II--P, show strong and narrow P--Cygni
features indicating very small expansion velocities ($\sim$1000 km s$^{-1}$ at 
the end of the plateau phase, see Fig. \ref{fig1}). 
In Pastorello et al. \cite{pasto03} and Zampieri et al. \cite{zamp03} 
other similar SNe were discussed (SNe 1994N, 1999br, 1999eu
and 2001dc).
The database has been recently enriched by the discovery of a well representative
event, SN~2003Z, extensively monitored at TNG\footnote{program TAC$\_$48}.
This SN provides a very good example of the spectro--photometric evolution of
low--luminosity SNe (see Fig. \ref{fig3}).
The SN was observed both in spectroscopy
and photometry during the photospheric phase, and observations during
the nebular phase, useful to estimate the $^{56}$Ni mass, are still in progress.
Preliminary late--time photometry suggests that SN~2003Z ejected 0.006 M$_\odot$
of $^{56}$Ni. We suggest that low--luminosity events may occur at an intrinsically high frequency.
\vspace{-.6cm}

\begin{figure}
\centering{\vspace{-0.2cm}
\includegraphics[height=6.4cm,width=9.5cm]{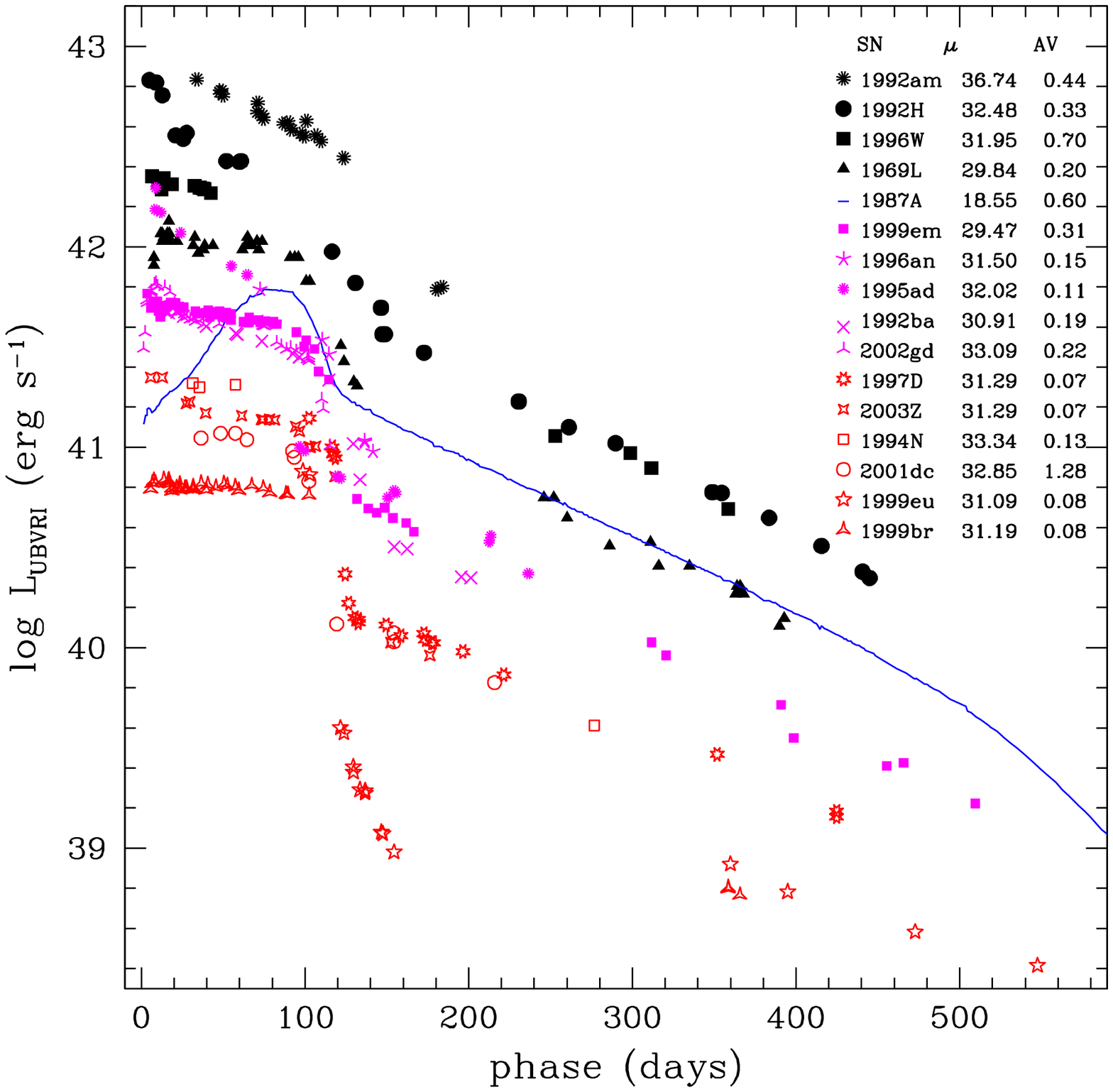}
\vspace{-0.25cm}
\includegraphics[height=6.4cm,width=9.15cm]{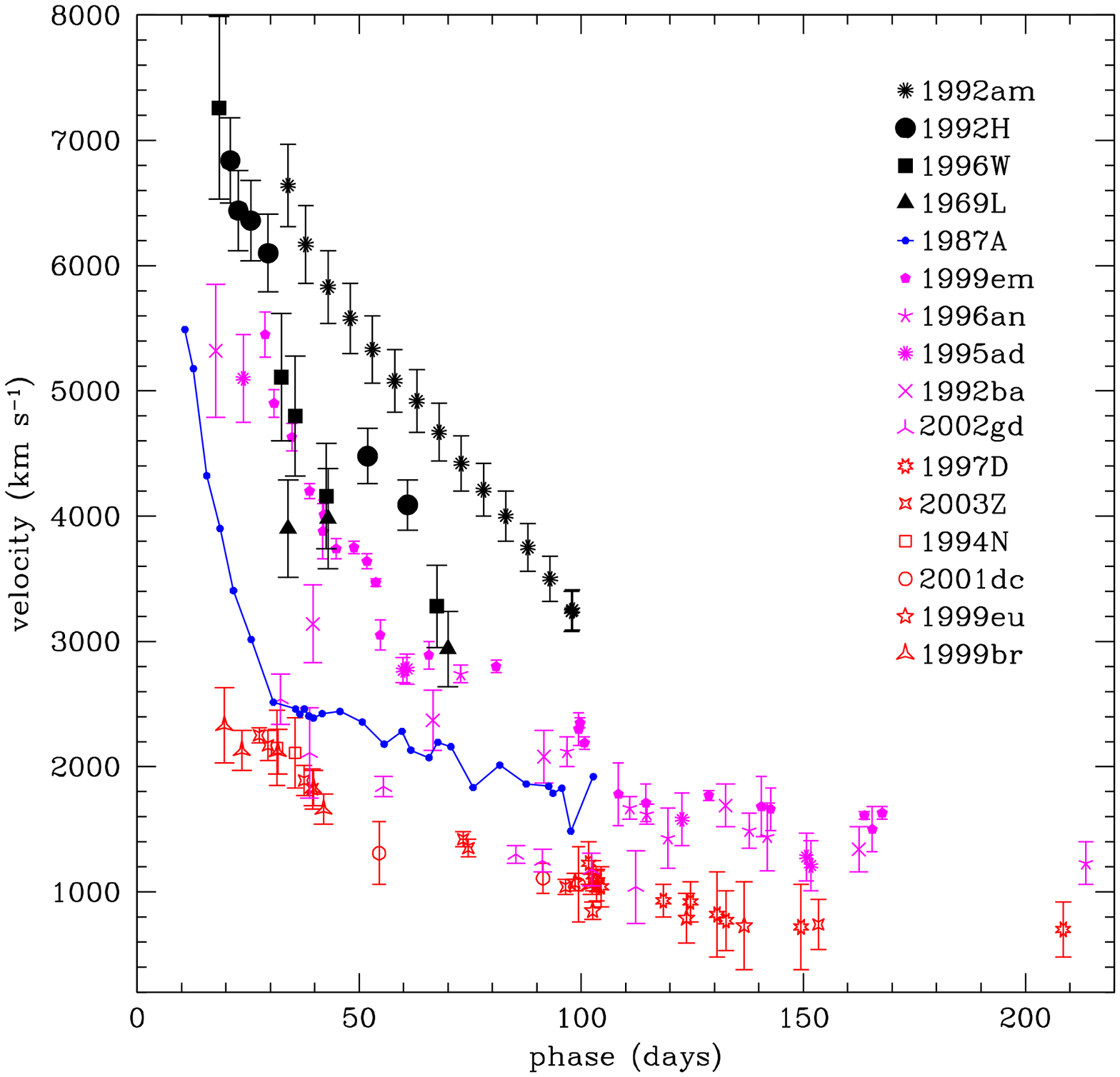}
\caption{Top: Luminosity evolution of our selected sample of SNe II--P. 
Bottom: expansion velocities obtained from the blueshift of the minimum of Sc II
lines. Adopted colours are:
red for faint 1997D--like SNe,
magenta for intermediate--luminosity SNe, black for normal and high luminosity events;
blue (solid line) is SN~1987A.}
\label{fig4}     }  
\end{figure}

The observed properties of the faint CC--SNe are consistent with 
very small ejected $^{56}$Ni mass ($<$ 10$^{-3}$ M$_\odot$)
and low explosion energy ($\ll$ 10$^{51}$ erg, \cite{zamp03}).
This suggests high--mass progenitors (M$_{MS} >$ 20--25 M$_\odot$)
for which significant fall--back might have occurred (\cite{zamp03}
and Zampieri et al., these Proceedings).

\vspace{-0.2cm}
\subsection{Normal SNe II--P}
The sample contains also a number of ``normal''
and high luminosity events covering a large range of physical properties
(e.g. Ni mass, explosion energy, ejected mass). 
Even if we observe a large spread both in luminosity
(2--20 $\times$ 10$^{42}$ erg s$^{-1}$) and in expansion velocity
(3000--5000 km s$^{-1}$ at the beginning of recombination), these SNe
never show the extreme properties of SN~1997D and other faint events.
Zampieri et al. (these Proceedings) suggest that the ejected envelope mass is 
in the range 12--26 M$_\odot$, with no definite tendency to vary with the
other SN parameters.

Peculiar is the case of SN~2002gd, well observed during the plateau, than
lost behind the sun. The plateau luminosity is relatively high, but the expansion velocity
deduced from the P--Cygni minima of spectral lines is small, close to that of faintest SNe II--P.
This SN was recently observed and our preliminary photometry suggests an unusually strong
post--plateau luminosity decrease. We may explain it with a very low amount of $^{56}$Ni ejected ($<$ 10$^{-3}$M$\odot$). 
Or, alternatively, dust formation into the ejecta might absorb the light at optical wavelengths,
leading us to underestimate the $^{56}$Ni mass. Because of its peculiar behaviour, other
late--time observations are required to better understand this event and before that any systematic 
analysis of its properties can be performed.
\vspace{-0.3cm}

\section{The heterogeneous family of SNe II--P}

A comparison among the pseudo--bolometric light curves for the SNe II--P of our sample
is shown in Fig. \ref{fig4}. The light curves appear to be heterogeneous in shape and
luminosity at all epochs. In particular
the exponential tails are powered by very different amounts
of $^{56}$Co (0.002--0.3 M$_\odot$). 
It's remarkable that the low--luminosity SNe are fainter at all stages
than all other SNe shown in Fig. \ref{fig4}. 

Also the evolution of the expansion velocity, obtained measuring the blueshift of the minima of the Fe II lines 
(see Fig. \ref{fig4}) shows a large spread at all epochs,
ranging from 3300 km s$^{-1}$ for SN 1992am \cite{schm94} 
to about 1000 km s$^{-1}$ for SN~1999br at $\sim$100 days after explosion. 
A similar spread is present also in the evolution of the continuum temperature.
This suggests, in accordance with \cite{hamu03}, that plateau luminosity, Ni mass, continuum 
temperature, expansion velocity and explosion energy
are correlated, from the high values of the luminous SNe 1992H and 1992am to the exceptionally
small ones for the faint SNe (see discussion in Zampieri et al., these Proceedings).
\vspace{-0.3cm}
%
%

%

\begin{thebibliography}{99.}
%
%
%
%
%
\bibitem{bene01} S. Benetti, M. Turatto, S. Balberg et al: MNRAS
\textbf{322}, 361 (2001)
\bibitem{ciat71} F. Ciatti, L. Rosino, F. Bertola: 
As. Contr. \textbf{255} (1971)
\bibitem{cloc96} A. Clocchiatti, M. M. Phillips, J. Spyromilio, B. Leibundgut: 
AJ \textbf{111}, 1286 (1996)
\bibitem{abou03} A. Elmhamdi, I. J. Danziger, N. N. Chugai et al.: MNRAS
\textbf{338}, 939 (2003)
\bibitem{hamu01} M. Hamuy: Type II Supernovae as Distance Indicators. 
PhD Thesis, University of Arizona (2001)
\bibitem{hamu03} M. Hamuy: ApJ \textbf{534}, 905 (2003)
\bibitem{leon02} D. C. Leonard, A. V. Filippenko, E. L. Gates et al., PASP,
\textbf{114}, 35 (2002)
\bibitem{pasto03} A. Pastorello, L. Zampieri, M. Turatto et al: MNRAS
\textbf{in press}, (2003)
\bibitem{pasto03tesi} A. Pastorello: Hydrogen 
Rich Core--Collapse Supernovae. PhD Thesis, University of Padova (2003)
\bibitem{nady03} D. K. Nadyozhin: MNRAS \textbf{in press}, (2003)
\bibitem{schl98} D. J. Schlegel, D. P. Finkbeiner, M. Davis et al: ApJ
\textbf{500}, 525 (1998)
\bibitem{schm94} B. Schmidt, R. P. Kirshner, R. G. Eastman et al: AJ
\textbf{107}, 1444 (1994)
\bibitem{tura98} M. Turatto, P. A. Mazzali, T. R. Young et al: ApJ
\textbf{498}, 129 (1998)
\bibitem{zamp03} L. Zampieri, A. Pastorello, M. Turatto et al: MNRAS
\textbf{338}, 711 (2003)

\end{thebibliography}
%



\printindex
\end{document}